# Single Pole-To-Earth Fault Detection and Location on the Tehran Railway System Using ICA and PSO Trained Neural Network

Masoud Safarishaal,

Amirkabir University Technology, Tehran, Iran

masoud.safarishaal66@gmail.com,

**Abstract:** In a railroad feeding system, detecting a location of pole to earth faults is important for safe operation of the system. The goal of this paper is to use a combination of the evolutionary algorithm and neural networks to increase the accuracy of single pole-to-earth fault detection and location on Tehran railroad power supply system. Accordingly, Imperialist Competitive Algorithm (ICA) and Particle Swarm Optimization (PSO) are used to train the neural network for enhancing learning process accuracy and the convergence. Owing to the nonlinearity of system, the fault detection is an ideal application for the proposed method where 600 Hz harmonic ripple method is used in this paper for fault detection. The substations were simulated by considering various situations in feeding the circuit, the transformer and the silicon rectifier has been developed by typical Tehran metro parameters. Required data for the network learning the process have been gathered from simulation results. 600Hz components value will change with the change of the location of single pole to earth fault. Therefore, 600Hz components are used as inputs of the neural network when fault location is the output of the network system. The simulation results show that the fault location can be accurately predicted in proposed methods.

# 1. INTRODUCTION

Nowadays, Transports on rail are increasing and major railroad infrastructure investments are expected. In a railroad feeding system, detecting a location of pole-earth faults is important for safe operation of the system. There are various types of fault sin railroad power traction systems, such as pole to pole (PP) faults, single pole to earth (PE) fault and pole to earth to pole (PEP) fault. Between these, PE maybe is not as much as other dangerous but detection and location of it can enable early preventive measures before the faults escalate into more dangerous poles to pole earth faults [1]. the fault current may be less than a normal train load, also many factors such as changing in train position, track nonlinear impedance, nonlinear source voltage of the substation rectifier and etc, so these faults are difficult to detect. makes it difficult to get the fault location simply by calculations. Owing to the nonlinearity of the system, the fault detections, an ideal application for the proposed method where the 600 Hz harmonic ripple method is used in this paper for fault detection [2]. To deal with these nonlinear situations, in this paper two modern neural network-based methods are explored. Todays, the idea of combining ANNs and evolutionary algorithms has received a lot of attention [3-8]. ICA is inspired by the historical pattern of competition among the countries of imperialism. In fact, this algorithm is an open door to the world of mathematics with a completely human perspective. The algorithm initially starts from several countries; In fact, these countries are possible answers to the algorithm. [9-11, 15-16]. Also, Particle Swarm Optimization (PSO) is a possible optimization. An evolutionary and simulated algorithm that is inferred from the behavior of humans and animals. A special feature of this algorithm (PSO) is that it executes real numbers directly in a continuous space and also, unlike other algorithms, does not guarantee the existence of a definite answer. PSO requires a small number of parameters to be set, which is easy to implement and has certain memory characteristics. [12], which have been inspired by nature and implemented in the different optimization problem [13]. In this paper, the simulation system is a typical simulation of the Tehran railway system. In fact, the electrical characteristics by simulation are similar to the basic characteristics of the actual system. 600Hz components value will change with the change of the location of single pole to earth fault. Therefore, 600Hz components are used as inputs of a neural network when fault location is the output of the network system. After this introductory section 1, the rest of this paper is organized as follows: In section 2, modeling of the railroad, components are presented. In section 3,600Hh components are described. Neural networks trained by evolutionary algorithms which have been used in this paper, are presented in section 4. Also, in section 5, simulation results are given and discussed. These results demonstrate the validity of the method. Also, in this section the performance of the ICA algorithm is compared with the performance of the PSO and conventional neural network method. Finally, in section 6 the conclusions are presented.

## 2. MODELING OF RAILWAY COMPONENT

### 2.1 Feeding System

The electric power generated by the power station is carried to electric railroad substations by transmission line. Traction power supply is one of the most important parts of the transporting structure. The scheme used in Tehran railroad electrification systems is to supply traction systems directly using the fundamental frequency main power, ie, 60 Hz [14]. The transmission or sub-transmission voltages are then directly transformed by a power transformer (63k/20k) to rectifier voltage. The output voltage of the rectifier to the traction voltage is 750v [14].

### 2.2 12-Pulse Rectifier:

Electric power must be converted by power converters so that it can flow between the public grid and the railroad grid. Because a three-phase power supply is used in Tehran Railway, a rectifier must be used to provide the required direct current. A 12-pulse system which is two sets of 6-pulse rectifier used in order to reduce the harmonics. In fact, this rectifier gives unregulated voltage. This voltage contains AC component of 600Hz and its harmonics. [2], in this paper this component used for fault detection.

### 2.3 Transformer

There are a supply transformer for each rectifier which provides the medium voltage for them needed based on its load. Therefore, transformer have two secondary windings, connected in the delta and the star at 600 Vac. The transformer is modeled simply with leakage reactance and winding resistance. Also, in Tehran's electrified railroad traction power supply system, the vector group is Yy0d1 [14]. The circuit of Fig.1 shows a parallel connected 12-pulse rectifier and its supply transformer.

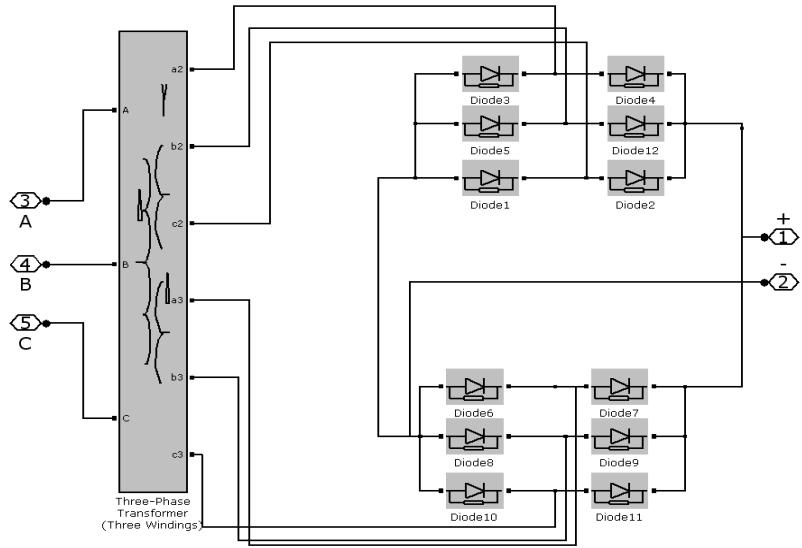

Fig.1. Tarsnsformer and 12 pulse rectifire

## 2.4 Traction

Traction system plays a very important role in railroad installations. Today, electric traction systems have been developed for both DC and AC power supply in railways. In Tehran railroad a NO.ZQ-132 DC motor is used as traction in this work the parameters of this motor have been used for simulation [14].

## 2.5 Total model

Fig.2 shows the total model of railroad system, as shown in it, bleed resistors of 100 and 200 Ohm are used both to define normal voltage with respect to earth, this resistor used to set the positive and negative rail normal voltage. X is the distance between the substation and fault location, also train location is assumed to be constant, in any simulation the value of X has been changed and the 600 Hz voltage amplitude is determined from this simulation result. R1 is the fault resistance that assumed to be 100 Ohm [1].

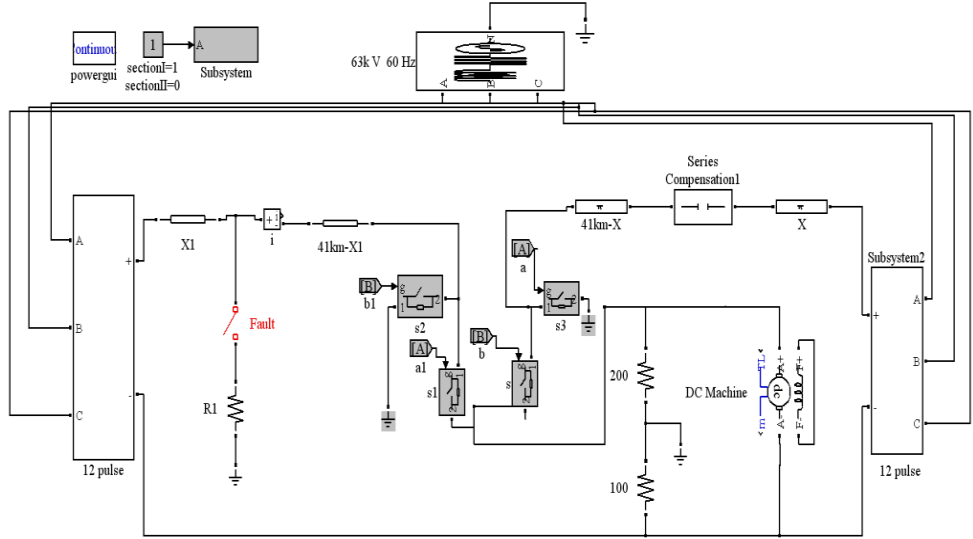

Fig.2. Total model of fault and railway

### 3. 600Hz COMPONENTS CIRCUIT

By using this circuit Vp, Vn, Ip and In can be measured. Accordingly, when other parameters such as train location Lt, train impedanc Zt and rail impedance Zrl are known the fault location can by calculated as following equation [1].

$$L_f = \frac{R_p(I_d - I_p) + R_n(I_d - I_n) - Z_t I_n - 2L_t Z_{rl} I_n}{Z_{rl}(I_p - I_n)} \quad (1)$$

$$= \frac{V_p - V_n - Z_t I_n - 2L_t Z_{rl} I_n}{Z_{rl}(I_p - I_n)} \quad (2)$$

However, the equations are different when Lt<Lf. Obtaining of the fault location by calculations, is hard because of many factors such as changing in train position, track nonlinear impedance, nonlinear source voltage of the substation rectifier etc[1]. To deal with these nonlinear situations, in this paper two modern neural network-based method are explored. Fig. 3 shows 600Hz component circuit with train position beyod the fault location [1].

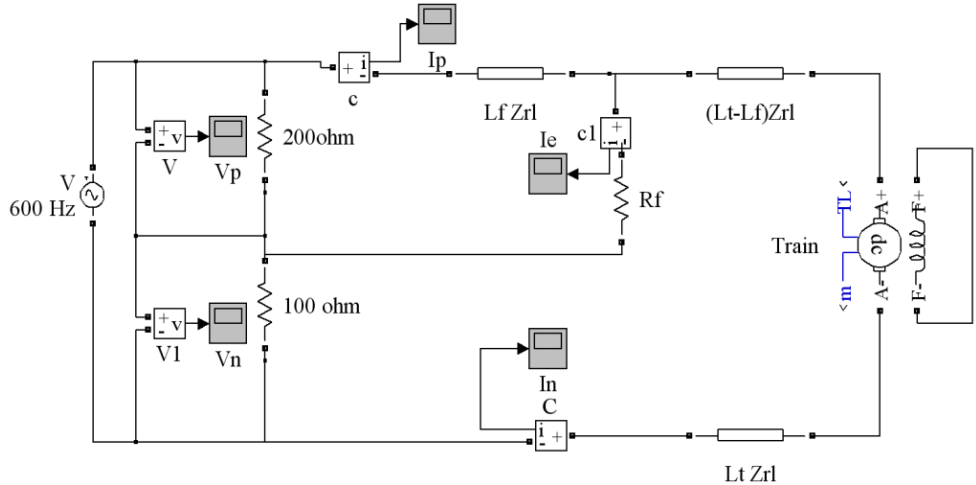

Fig.3. 600Hz Component circuit with train position beyod the fault location.

## 4. NEURAL NETWORK TRAINED BY EVOLUTIONARY ALGORITHMS

Using traditional neural network design methods is not very accurate. In this paper, we intend to present a different method. In this method, we first use the ICA and PSO to train neural networks, then trained neural networks are used to find the location of the fault.

### 4.1 Imperialist Competitive Algorithm

The Imperialist Competitive Algorithm (ICA) is one of the newest evolutionary optimization algorithms. This algorithm, as its name implies, is based on modeling the socio-political process of the colonial phenomenon. Therefore, it is a new algorithm and can compete with other evolutionary algorithms such as genetic algorithms and ant colony algorithms, etc. In terms of application, it has been used to solve many problems in the field of optimization, including in electrical engineering, computer, industry, mechanics, economics, management, etc. The reason for the high popularity of this algorithm, along with its high efficiency, is mostly due to its innovative, new and attractive aspect for optimization experts. Fig. 4 shows the flowchart of the ICA. In this paper, a population of 30 countries consisting of 3 empires and 27 colonies are used.

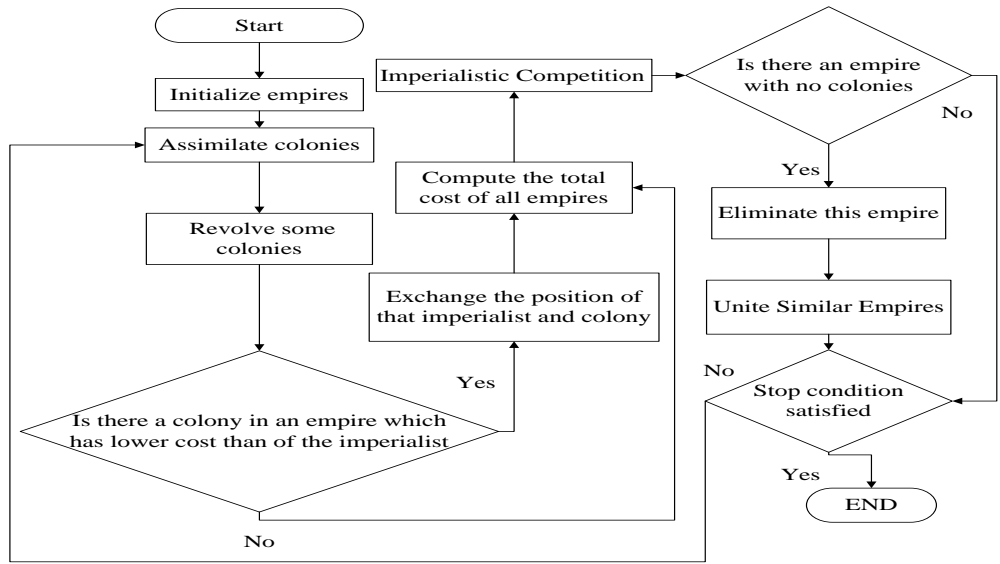

Fig.4. Flowchart of the Imperialist Competitive Algorithm

## 4.2  Particle Swarm Optimization

The idea of the Particle Swarm Optimization algorithm was first proposed by Kennedy and Eberhart [12]. PSO is a nature-inspired, iterative evolutionary computational algorithm. The source of inspiration for this algorithm is the social behavior of animals, such as the collective movement of birds and fish. As the name implies, it is a particle-based optimization method. The basis for the development of the PSO algorithm is that the possible solutions to an optimization problem are considered as birds without volume and qualitative properties, which are referred to as particles, these birds fly in an n-dimensional space and their path in space. They change the search based on their past experiences and those of their neighbors. The PSO algorithm has a simple structure and at the same time has acceptable performance. Therefore, due to the capability of the PSO algorithm in finding the optimal global solution with a very high probability and high convergence rate, this algorithm has been used to train neural networks. Fig. 5 shows the flowchart of the PSO. In this case of study, 200 particles and 400 iterations were employed.

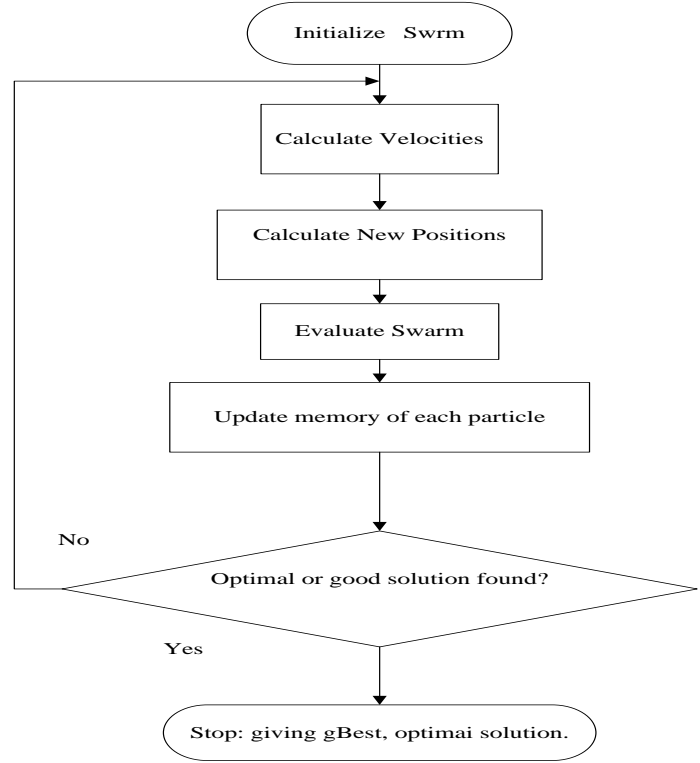

Fig.5. Flow chart for Particle Swarm optimization

## 5. RESULTS

In this section two proposed estimator systems, compared with conventional feed forward neural network and with each other Fig. 6 (a) and (b) show ICA trained NN output vs. real output for train data and test data respectively, also Fig. 7 (a) and (b) show PSO trained neural network output vs. real output for training and test data. Accordingly, Fig. 8 (a) and (b) show output vs. real output for training and test data. In these figures the correlation of real output with proposed method output is displayed. In addition to correlation rate, in order to compare the accuracy of proposed methods, it is very important to select a reasonable metric to evaluate the results. Hence, the following Mean Absolute Percentage Error (MAPE) and Root Mean Square Error (RMSE) are used here for after-the-fact error analysis:

$$\varepsilon = \frac{1}{N}\sum_{i=1}^{N}\frac{|x_t - x_f|}{x_t}*100 \qquad (6)$$

$$\partial = \sqrt{\frac{1}{N}\sum_{i=1}^{N}(x_t - x_f)^2} \qquad (7)$$

Where Xt is the actual load and Xf is the forecasted data. MAPE is a percentage of error while RMSE is a number indicating the model bias. MAPE, RMSE correlation value for all mentioned method measured for training and testing data as shown in Table. I. As the results show, accuracy of the model performance was acceptable in both proposed methods, although ICA NN's results were slightly more accurate.

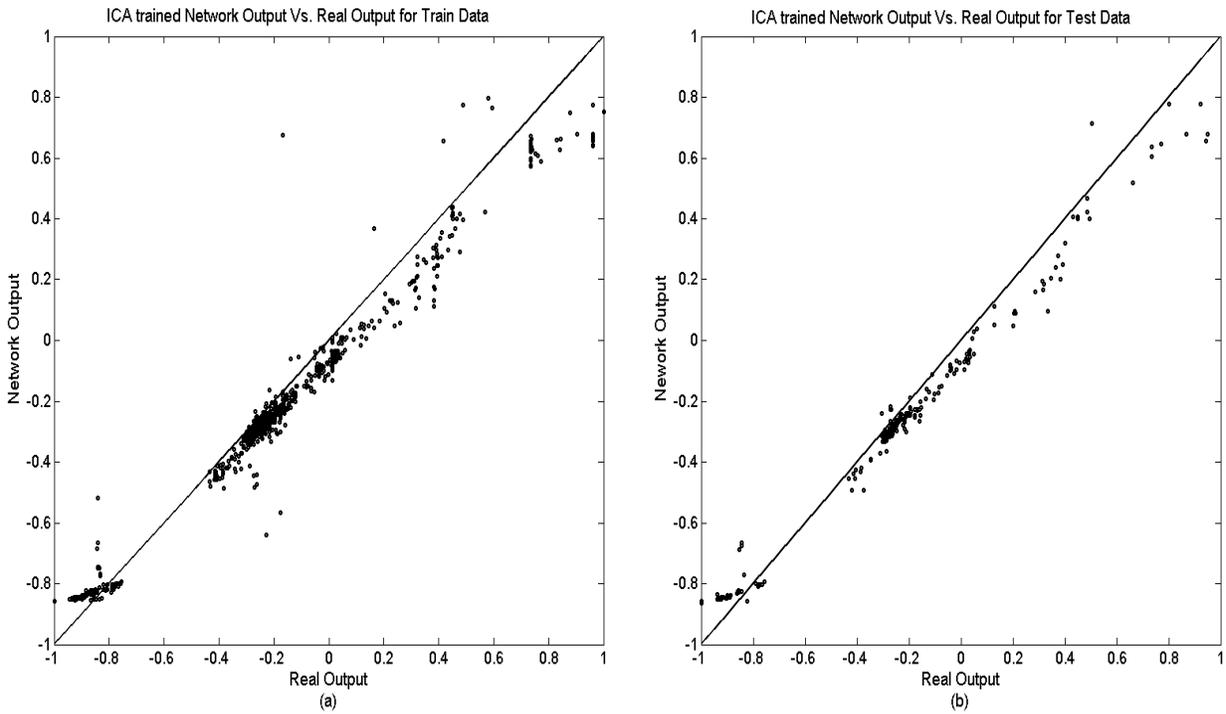

Fig.6. ICA trained Neural Network output vs. real output for a) train data b) test data

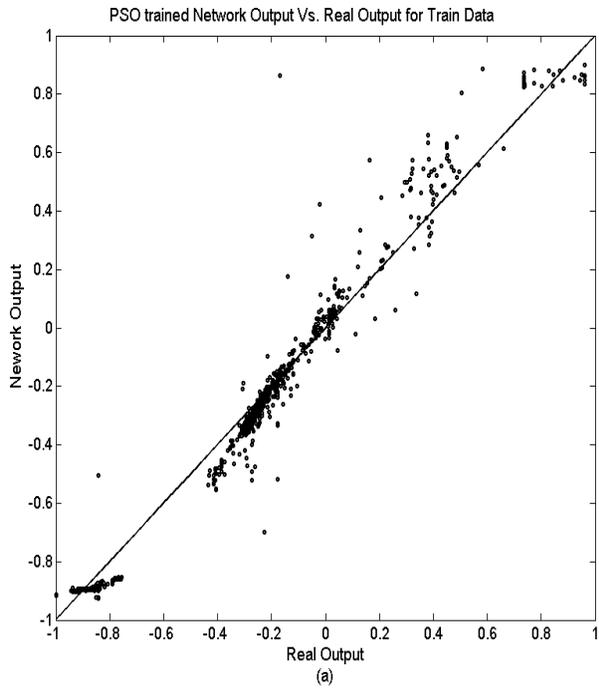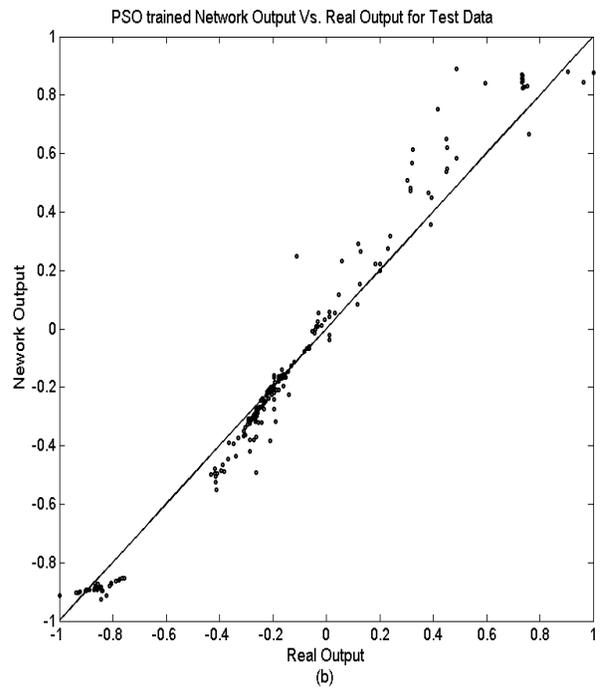

Fig.7. PSO trained Neural Network output vs. real output for a) train data b) test data

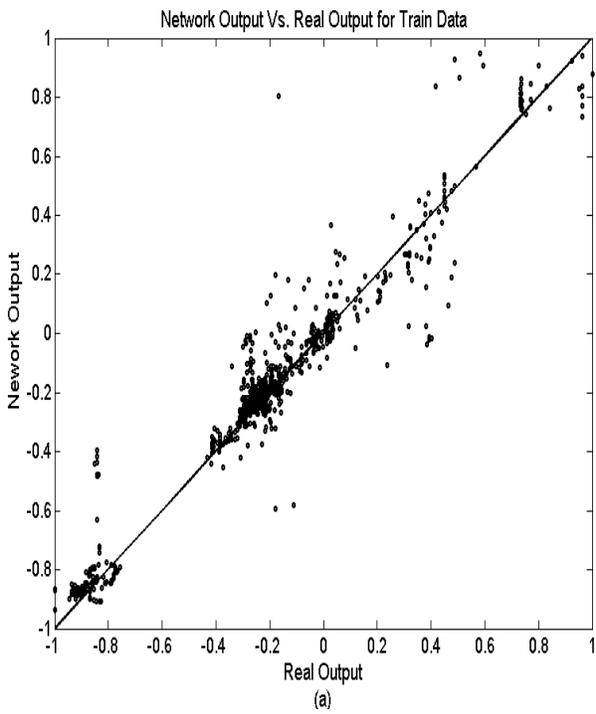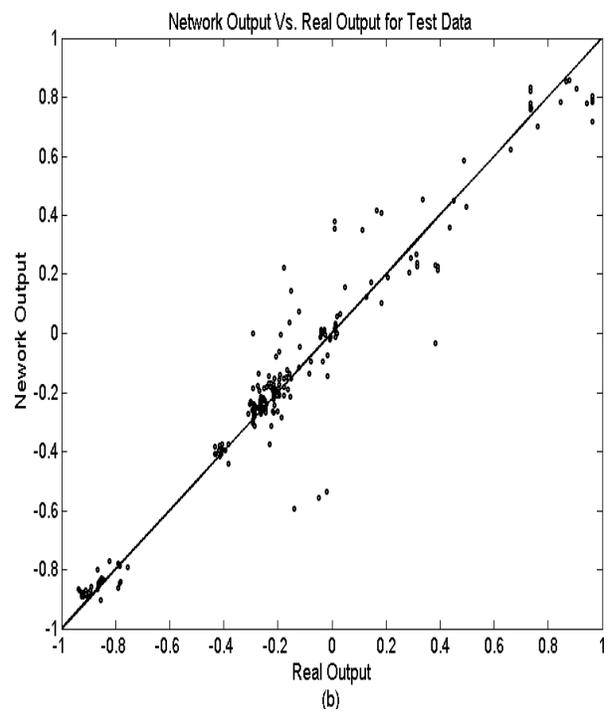

Fig.8. Feed Forward Neural Network output vs. real output for a) train data b) test data

Table. I. Comparison of proposed method with conventional Neural Network

|  | Training data | | | Testing Data | | |
| --- | --- | --- | --- | --- | --- | --- |
|  | MAPE | RMSE | Correlation | MAPE | RMSE | Correlation |
| ICA trained NN | 0.0061 | 0.031 | 0.9885 | 0.0055 | 0.026 | 0.9927 |
| PSO trained NN | 0.0060 | 0.042 | 0.9840 | 0.0067 | 0.038 | 0.9883 |
| Neural Network | 0.0099 | 0.089 | 0.9679 | 0.0108 | 0.085 | 0.9718 |

## 6. CONCLUSION

A typical simulation of the Tehran railroad system is developed in MATLAB environment. Single pole to earth fault detection using 600Hz ripple components by three neural networks models, namely ICA trained neural network, PSO trained neural network and conventional feed forward neural network are presented. The result shows that detection can be improved by combination of the mentioned evolutionary algorithm by neural network. Also, the ICA trained the neural network to give a better result than other methods.